\documentclass[12pt]{article}
\usepackage{epsfig}
\usepackage{multirow}

\newcommand{\mysection}{\setcounter{equation}{0}\section}

\def\beq{\begin{equation}}
\def\eeq{\end{equation}}
\def\beqa{\begin{eqnarray}}
\def\eeqa{\end{eqnarray}}
 
\newlength{\dinwidth} \newlength{\dinmargin}
\begin{document}

\begin{center}
{\Large \bf The top quark forward-backward asymmetry at approximate N$^3$LO}
\end{center}
\vspace{2mm}
\begin{center}
{\large Nikolaos Kidonakis}\\
\vspace{2mm}
{\it Department of Physics, Kennesaw State University,\\
Kennesaw, GA 30144, USA}
\end{center}
 
\begin{abstract}
I calculate the top quark forward-backward asymmetry at the Tevatron in 
both the laboratory frame and the $t{\bar t}$ rest frame. 
I show that soft-gluon corrections are the dominant contribution to the 
asymmetry and closely approximate exact results through 
next-to-next-to-leading order (NNLO). 
I present a calculation of the asymmetry including approximate next-to-next-to-next-to-leading-order (N$^3$LO) soft-gluon contributions from next-to-next-to-leading-logarithm (NNLL) resummation as well as electroweak corrections. 
Thus approximate N$^3$LO  (aN$^3$LO) results are obtained, which 
significantly enhance and improve previous NNLO results. 
The theoretical aN$^3$LO result for the 
top quark forward-backward asymmetry at the Tevatron in the laboratory frame 
is ($6.8 \pm 0.3$)\%,  and in the $t{\bar t}$ rest frame it is 
($10.0 \pm 0.6$)\% which is in excellent agreement with recent Tevatron data.
\end{abstract}
 
\mysection{Introduction}

Top quark production has been studied extensively at the Tevatron and the LHC. 
The soft-gluon corrections are now known approximately to  
next-to-next-to-next-to-leading-order (N$^3$LO) \cite{NKaNNNLO,NKpty}, 
based on the next-to-next-to-leading-logarithm (NNLL) two-loop 
soft-gluon resummation formalism of \cite{NKpt}, which allows the calculation 
of approximate N$^3$LO (aN$^3$LO) results for the total cross 
section \cite{NKaNNNLO} and the 
transverse momentum and rapidity distributions \cite{NKpty}. 
The agreement between theory and experimental results from the LHC and 
the Tevatron is excellent for all available energies for both the total 
and differential cross sections.

A quantity of particular interest is the top quark forward-backward asymmetry,
$A_{\rm FB}$, which is closely related to the rapidity distribution of the top 
quark. The definition of the asymmetry is     
\beq
A_{\rm FB} =\frac{\sigma(y_t>0)-\sigma(y_t<0)}{\sigma(y_t>0)+\sigma(y_t<0)}\equiv 
\frac{\Delta\sigma}{\sigma} \, ,
\label{AFB}
\eeq
where $\sigma$ denotes cross sections and $y_{t}$ is the top-quark rapidity in
a given Lorentz frame, which may be the $p{\bar p}$ laboratory frame or the 
$t{\bar t}$ rest frame (in the latter frame the asymmetry is equivalently
defined in terms of $\Delta y=y_t-y_{\bar t}$, with $y_{\bar t}$ the antitop 
rapidity; see  Ref. \cite{WBZS} for more details on various definitions and 
further references). 
At a $p\bar p$ collider such as the Tevatron, charge conjugation invariance in
QCD implies that $\sigma(y_{\bar t}>0)=\sigma(y_{ t}<0)$, so the forward-backward
asymmetry is equivalent to the charge asymmetry.
The $t{\bar t}$ rest frame asymmetry is larger than 
the laboratory frame asymmetry by about fifty percent due to kinematics. 

Early measurements of $A_{\rm FB}$ were performed 
by CDF \cite{CDFlab} and D0 \cite{D0lab} and 
returned values substantially larger than standard model predictions.
Recent measurements of $A_{\rm FB}$ in the $t{\bar t}$ rest frame with more 
data and precision have been reported in \cite{CDF,D0}.
Thus, it is important to perform the most accurate calculation for the 
asymmetry in order to have confidence in the theoretical prediction while 
looking for hints of new physics. 

The QCD corrections to the asymmetry were calculated with 
next-to-leading-logarithm (NLL) resummation of soft-gluon contributions 
in the moment-space formalism in \cite{ASV}. 
This was later extended to NNLL accuracy in \cite{NKy} 
where approximate next-to-next-to-leading-order (NNLO) results were obtained.
Results at NNLL were also derived using Soft-Collinear Effective Theory 
(SCET) in \cite{AFNPY}. An approach using the principle of maximum 
conformality was used in \cite{SBXW} to calculate the asymmetry.
Recently the complete NNLO QCD contribution to the asymmetry has appeared 
in \cite{CFM}.

A variety of resummation formalisms and methods have been proposed in the 
literature for top quark production over the past twenty years (see the 
discussions in \cite{NKpt,NKPRD64,NKBP,NKHQ13} and references therein). 
These differences include whether resummation is performed for the total cross section or for double-differential cross sections (i.e. absolute vs partonic threshold); the inclusion and motivation for various subleading terms and whether or not they are warranted; the use of damping factors away from threshold; whether resummation is performed in moment-space perturbative QCD vs SCET resummation; the use of prescriptions (and their validity) in resummed results vs performing prescription-independent fixed-order expansions at NNLO or N$^3$LO; the choice of kinematics, single-particle-inclusive (1PI) vs pair-invariant-mass (PIM), for double-differential resummations, and various choices for writing the analytical terms via partonic threshold relations of kinematical variables; etc. More details and references may be found in \cite{NKpt,NKPRD64,NKBP,NKHQ13}. In most recent formalisms (see Refs. \cite{NKpt,NKy,NKBP,NKHQ13,ALLMUW,BFKS}) the soft-gluon corrections were found to be large and dominant. 

Most resummation formalisms for top production use methods applicable only to calculations of total cross sections. Only two formalisms have been developed at NNLL for double-differential cross sections and applied to the asymmetry, 
one using moment-space QCD resummation \cite{NKy} 
and the other using SCET \cite{AFNPY}.
The differences between the moment-space QCD and the SCET resummation 
approaches in top production were discussed in the review paper \cite{NKBP}, 
where results for the asymmetry in different formalisms were shown. In the 
formalism of \cite{NKy} the approximate NNLO contributions to the asymmetry 
were found to be significant while in the approach of \cite{AFNPY} they were 
rather small. This discrepancy was not unexpected since the overall 
contributions from soft-gluon resummation to the total cross section and 
differential distributions were also quite different in the two approaches. 
There are several differences between the two resummations, as described 
in \cite{NKBP}, but the numerically prominent one is the inclusion of certain 
terms in \cite{AFNPY} that arise in the SCET approximation and are not strictly 
of soft gluon origin. Those SCET terms severely depress the results in 
\cite{AFNPY} relative to \cite{NKy}. As we will show in Section 2, the result 
with large  
approximate NNLO corrections in \cite{NKy} precisely predicted the exact NNLO 
asymmetry; the results in \cite{AFNPY} are much smaller, with tiny  
corrections beyond next-to-leading order (NLO).   
 
The fact that the resummation formalism used in \cite{NKpt,NKy} predicts 
the exact NNLO cross section better than all other approaches, to better than 
one per mil accuracy for the central value at the Tevatron, as shown in \cite{NKHQ13}, had already strongly indicated that the results for the asymmetry in \cite{NKy} would be very close to the exact NNLO result before the latter was known. This is indeed the case as we discuss in the next section. 
We note that there are strong theoretical reasons for the choices we have made in our approach. Formally, the resummation uses renormalization group methods to fully determine terms at a given logarithmic accuracy (NNLL in this case). Numerically, subleading terms have to be treated carefully to approximate and predict the corrections correctly. At NLO the total cross sections as well as the transverse momentum and rapidity distributions at all Tevatron and LHC energies were reproduced nearly exactly with our method. It was also shown that with our method the differences between approximate NNLO results in 1PI and PIM kinematics are minimized, which indicated that any missing terms were numerically small \cite{NKRV}. This fact together with the correct description at NLO was a very strong indicator that our NNLO predictions would be successful and, not surprisingly, this has indeed been the case, to very high accuracy, which validates our method and the theoretical arguments in support of it. 

Parenthetically we note that it has been known for a while that  
the NLO expansion of the results from soft-gluon 
resummation for  the asymmetry reproduces the complete NLO result rather well 
(this was noted already in \cite{ASV} where NLL resummation was used).  
The extension to NNLL in \cite{NKy} has a similar effect in the NNLO expansion,
in predicting the exact NNLO results of \cite{CFM} very well.

The authors of \cite{CFM} also found a significant NNLO contribution, as 
expected from \cite{NKy}. However, the authors of \cite{CFM} 
stated erroneously that the effects of soft-gluon contributions are negligible.
As discussed above, and as will be shown in more detail in the next section, 
this is factually incorrect. The soft-gluon contributions have been known to 
be large and dominant in the total and differential top production cross 
sections for at least twenty years (for references see the review 
in \cite{NKBP}), and in the asymmetry in particular at least 
since 2011 \cite{NKy}. In fact the NNLO enhancements to the asymmetry 
found in \cite{CFM} are very close to the predictions in \cite{NKy}. 
The accuracy, precision, reliability, and robustness of the resummation 
formalism that we use have been amply demonstrated and discussed in 
\cite{NKpt,NKHQ13} for total cross sections and differential distributions.  

We also note that while our resummation formalism uses the soft-gluon eikonal 
approximation for partonic threshold, the integrations are over all allowable 
gluon energies. Furthermore, partonic threshold is more general than absolute 
production threshold, i.e. the top quark is not necessarily produced at rest 
but may have arbitrary transverse momentum and rapidity as allowed by the 
kinematics. As a consequence, the effects of these approximate contributions to
the top-quark transverse momentum ($p_T$) distribution increase with $p_T$ at 
large $p_T$ since that is a region closer to partonic threshold (see e.g. Fig. 3 in the first paper of Ref. \cite{NKHQ13}). Thus, we find 
the arguments made in \cite{CFM} regarding the contributions of soft gluons to 
the transverse momentum distribution of the top-antitop pair 
to be invalid or irrelevant in our approach.

In this paper we will improve on the NNLO results for the asymmetry 
by further calculating N$^3$LO soft-gluon corrections, and including 
electroweak corrections \cite{WBZS,WBS10,WHDP,AMMT}, thus producing an 
aN$^3$LO estimate. We present the results for the aN$^3$LO asymmetry  
in both the laboratory frame and 
the $t{\bar t}$ rest frame.

\mysection{The top-quark forward-backward asymmetry at the Tevatron}

We write the perturbative QCD expansion for the top-antitop pair cross section 
as 
\beq
\sigma=\sum_{n=0}^{\infty} {\alpha_s}^{2+n} \, \sigma^{(n)}
\eeq
with $\alpha_s$ the strong coupling. At leading order the cross section 
is of order $\alpha_s^2$, the NLO corrections $\sigma^{(1)}$ are of order 
$\alpha_s^3$, etc. We include the complete $\sigma^{(1)}$ and $\sigma^{(2)}$ 
corrections and the approximate $\sigma^{(3)}$ corrections from NNLL 
resummation.

At each order in $\alpha_s$, the cross section includes plus distributions
of the form $[\ln^k(s_4/m_t^2)/s_4]_+$ where $s_4=s+t+u-2m_t^2$, 
with $s$, $t$, $u$ the usual kinematical variables and $m_t$ the top-quark 
mass. The variable $s_4$ measures distance from partonic threshold
and, for the $n$th-order corrections, 
the power of the logarithm, $k$, in the plus distributions can range from the 
leading value of $2n-1$ down to the lowest value of 0. 
Thus, at NLO the leading value of $k$ is 1, at NNLO it is 3, 
and at N$^3$LO it is 5. With NNLL resummation one can derive the 
coefficients of all powers of the logarithms at NLO and NNLO, but not at 
N$^3$LO. We use the inverse
transform of the moment-space logarithms of the Mellin moment to the
specified order, including constant $\zeta_i$ terms (see also discussion in 
\cite{NKBP}).
As discussed in \cite{NKaNNNLO}, at N$^3$LO the 
coefficients of the powers of the logarithms with $k=5,4,3,2$ can be 
fully determined, but those with $k=1,0$ are only partially known and 
include factorization and renormalization scale terms as well as  
terms arising from the inversion from moment space back to momentum space 
(see Ref. \cite{NKPRD73} for explicit expressions). 
The inversion terms, involving $\zeta_i$ 
constants, are known to dominate the subleading powers of the logarithms, 
and the impact of the remaining subleading terms is much smaller than the 
scale variation \cite{NKaNNNLO}.

The top quark forward-backward asymmetry is then written as
\beq
A_{\rm FB}=\frac{\Delta\sigma^{\rm EW}+\alpha_s^3 \, \Delta\sigma^{(1)}
+\alpha_s^4 \, \Delta\sigma^{(2)}+\alpha_s^5 \, \Delta\sigma^{(3)}+\cdots}
{\alpha_s^2 \, \sigma^{(0)}+\alpha_s^3 \, \sigma^{(1)}+\alpha_s^4 \, \sigma^{(2)}
+\alpha_s^5 \, \sigma^{(3)}+\cdots}
\label{AFBnoexp}
\eeq
where again $\Delta\sigma^{(n)}$ denotes the QCD asymmetric terms 
as in Eq. (\ref{AFB}), i.e. the 
difference of $n$th-order QCD
corrections integrated over positive and negative rapidities, and 
$\Delta\sigma^{\rm EW}$ denotes the asymmetric term from electroweak corrections. 

One may further reexpand the above ratio in the definition of $A_{\rm FB}$ in
powers of $\alpha_s$, thus obtaining the expression
\beqa
A_{\rm FB} &=& \frac{\Delta\sigma^{\rm EW}}{\alpha_s^2 \, \sigma^{(0)}} 
+\alpha_s \frac{\Delta\sigma^{(1)}}{\sigma^{(0)}}
-\frac{\Delta\sigma^{\rm EW} \sigma^{(1)}}{\alpha_s \, (\sigma^{(0)})^2}
+\alpha_s^2 \left[\frac{\Delta\sigma^{(2)}}{\sigma^{(0)}}
-\frac{\Delta\sigma^{(1)} \, \sigma^{(1)}}{(\sigma^{(0)})^2}\right]
\nonumber \\ &&
{}+\frac{\Delta\sigma^{\rm EW}}{(\sigma^{(0)})^3}
\left[(\sigma^{(1)})^2-\sigma^{(0)} \, \sigma^{(2)}\right]
+\alpha_s^3 \left[\frac{\Delta\sigma^{(3)}}{\sigma^{(0)}}
-\frac{\Delta \sigma^{(2)} \, \sigma^{(1)}}{(\sigma^{(0)})^2}
+\frac{\Delta \sigma^{(1)} (\sigma^{(1)})^2}{(\sigma^{(0)})^3}
-\frac{\Delta \sigma^{(1)} \, \sigma^{(2)}}{(\sigma^{(0)})^2}\right] 
\nonumber \\ &&
{}+\cdots
\label{AFBexp}
\eeqa

Explicit formulas and numerical values for the quantities $\Delta \sigma^{(i)}$ 
and $\sigma^{(i)}$ may be found in Refs. \cite{NKaNNNLO,WBZS,CFM,NKPRD73}.
In the following we will provide results for both definitions, 
Eq. (\ref{AFBnoexp}) and Eq. (\ref{AFBexp}), in both the 
laboratory and $t{\bar t}$ rest frames.

In the Standard Model the forward-backward asymmetry is due mainly to QCD 
effects. At leading order in perturbative QCD, the production channels 
$q{\bar q} \rightarrow t{\bar t}$ and $gg \rightarrow t{\bar t}$ 
are symmetric in rapidity, thus $A_{FB}$ vanishes at that order. 
Furthermore, the $gg$ channel remains 
symmetric at all orders in perturbative QCD. 
However an asymmetry arises in the $q{\bar q}$ channel 
(as well as a much smaller contribution from the the $qg$ channel) starting at NLO. 
Therefore by applying soft-gluon resummation we expect the $gg$ channel to 
remain symmetric, but we have contributions to the asymmetry from higher 
orders in the $q{\bar q}$ channel.

\subsection{$p{\bar p}$ laboratory frame}

We first provide results for the asymmetry 
in the laboratory frame. We use MSTW2008 \cite{MSTW}
parton distribution functions and a top quark mass $m_t=173.3$ GeV in the 
calculations.  
The central values refer to the choice of scale $\mu=m_t$,
and theoretical errors are estimated with scale variation by a factor of two. 

Using the NNLO approximate rapidity distributions it was found in \cite{NKy} 
that the approximate NNLO top quark forward-backward asymmetry 
in the laboratory frame at the Tevatron 
(from QCD effects only) is  5.2\%, using Eq. (\ref{AFBnoexp}). 
This is a large increase of thirty percent over the NLO asymmetry (and as 
we will see later, we find essentially the same percent increase in the 
$t{\bar t}$ rest frame). We also note that this NNLO / NLO ratio is very 
similar to the increase
that was found in \cite{CFM}, as we also discuss in the next subsection. 
We confirm the results of \cite{NKy} in this paper and we add the aN$^3$LO QCD  
corrections as well as electroweak corrections. Furthermore, 
we calculate the corresponding numbers using Eq. (\ref{AFBexp}).    

At aN$^3$LO we find that the asymmetry from QCD alone is ($5.6^{+0.3}_{-0.4}$)\% 
using Eq. (\ref{AFBnoexp}), and ($6.0 \pm 0.1$)\% using Eq. (\ref{AFBexp}).
The uncertainties indicated in these numbers are estimated by varying the 
scale $\mu$ between $m_t/2$ and $2m_t$.
If we include the electroweak corrections as well as the aN$^3$LO QCD 
corrections we find for the asymmetry the results ($6.4^{+0.5}_{-0.6}$)\% 
using Eq. (\ref{AFBnoexp}),  
and ($6.8 \pm 0.3$)\% using Eq. (\ref{AFBexp}). The aN$^3$LO results with and 
without electroweak corrections are tabulated in Table 1.

We also note that the differences between using Eq. (\ref{AFBnoexp}) 
and Eq. (\ref{AFBexp}) decrease as higher orders are included 
in the calculation of $A_{FB}$.

\begin{table}[htb]
\begin{center}
\begin{tabular}{|c|c|c|} \hline
\multicolumn{3}{|c|}{Top-quark asymmetry at the Tevatron} \\ \hline
aN$^3$LO $A_{FB}$ \% & $p{\bar p}$ frame & $t{\bar t}$ frame \\ \hline
QCD only Eq. (\ref{AFBnoexp}) & $5.6^{+0.3}_{-0.4}$  & $8.1^{+0.4}_{-0.6}$ \\ \hline 
QCD only Eq. (\ref{AFBexp}) & $6.0 \pm 0.1$ & $8.7 \pm 0.2$  \\ \hline 
QCD+EW Eq. (\ref{AFBnoexp}) & $6.4^{+0.5}_{-0.6}$ & $9.4^{+0.7}_{-0.9}$ \\ \hline
QCD+EW Eq. (\ref{AFBexp}) & $6.8 \pm 0.3$ & $10.0 \pm 0.6$ \\ \hline
\end{tabular}
\caption[]{The top-quark forward-backward asymmetry in $p{\bar p}$ collisions 
at the Tevatron in aN$^3$LO QCD with and without electroweak (EW) corrections, 
using $\sqrt{S}=1.96$ TeV and $\mu=m_t=173.3$ GeV. Results are shown in the 
$p {\bar p}$ (laboratory) frame and the $t{\bar t}$ rest frame.}
\label{table}
\end{center}
\end{table}

\subsection{${t \bar t}$ rest frame}

In the ${t \bar t}$ rest frame the overall asymmetry is around 
fifty percent larger than 
in the laboratory frame, due to kinematics, but the overall 
ratios of contribution levels at different orders in QCD are similar.  
As in the laboratory frame, in the ${t \bar t}$ rest frame the increase from 
NNLO contributions from QCD is around thirty percent on top of the NLO 
asymmetry, using Eq. (\ref{AFBnoexp}). More precisely, the ratio of the aNNLO 
to the NLO asymmetry is 1.29, which is very close to the value of 1.27 for the 
NNLO over NLO ratio that was found in \cite{CFM}. Thus, approximate NNLO 
results for the asymmetry are nearly identical to exact NNLO results. 
Our aNNLO result is ($7.6^{+0.4}_{-0.9}$)\% which is only one percent higher than 
the exact NNLO result in \cite{CFM} and has very similar scale uncertainty. 
The NNLO result and the increase over NLO is practically the same whether one 
uses the exact NNLO corrections \cite{CFM} or the NNLO soft-gluon 
contributions, as before. As we will show in the figures in the next 
subsection, this agreement holds not just for the total asymmetry but also in 
separate rapidity bins.
 
There is also good agreement between exact and approximate results if one uses
Eq. (\ref{AFBexp}) instead. We find an aNNLO/NLO ratio of 1.14 vs the 
NNLO/NLO ratio of 1.13 in \cite{CFM}. Specifically, our value for the 
QCD aNNLO asymmetry is ($8.4 \pm 0.2$)\%, which again is only one percent 
higher than the exact NNLO value in \cite{CFM}.
All these results again confirm the fact that the overwhelming QCD contribution
to the asymmetry can be derived from the soft-gluon approximation of 
Ref. \cite{NKpt}, which again is fully consistent with similar findings for the 
total cross section and the transverse momentum and rapidity distributions of 
the top quark \cite{NKaNNNLO,NKpty,NKpt,NKy,NKHQ13}.
   
At aN$^3$LO we find that the asymmetry from QCD alone is ($8.1^{+0.4}_{-0.6}$)\% 
using Eq. (\ref{AFBnoexp}), and ($8.7 \pm 0.2$)\% using Eq. (\ref{AFBexp}).
The uncertainties indicated in these numbers are estimated by varying the 
scale $\mu$ between $m_t/2$ and $2m_t$, as before.
If we include the electroweak corrections as well as the aN$^3$LO QCD 
corrections, we find for the asymmetry the results ($9.4^{+0.7}_{-0.9}$)\% 
using Eq. (\ref{AFBnoexp}), and ($10.0 \pm 0.6$)\% using Eq. (\ref{AFBexp}). 
The aN$^3$LO results with and without 
electroweak corrections are again tabulated in Table 1.

We note that the N$^3$LO soft-gluon corrections provide an 8\% increase over 
the NNLO QCD result for the asymmetry using Eq. (\ref{AFBnoexp}). The increase 
over NNLO becomes 5\% if instead Eq. (\ref{AFBexp}) is used.
Thus, the increase from the aN$^3$LO corrections is very substantial; in 
addition, the scale dependence is reduced at aN$^3$LO relative to NNLO.

The best estimate for $A_{\rm FB}$ in the ${t \bar t}$ rest frame 
is ($10.0 \pm 0.6$)\% which is in excellent 
agreement with the D0 result of ($10.6 \pm 3.0$)\% \cite{D0} and the 
CDF result of ($16.4 \pm 4.7$)\% \cite{CDF}.

\subsection{Differential $A_{\rm FB}$}

In this subsection we present the differential forward-backward asymmetry 
defined by 
\beq
A^{\rm bin}_{\rm FB} =\frac{\sigma^+_{\rm bin}(\Delta y)-\sigma^-_{\rm bin}(\Delta y)}
{\sigma^+_{\rm bin}(\Delta y)+\sigma^-_{\rm bin}(\Delta y)} 
\label{AFBbin}
\eeq
with $\Delta y=y_t-y_{\bar t}$ and where the bin determines the range of 
$\Delta y$. In the figures below we present results for four $|\Delta y|$ bins 
of width 0.5 over the range from 0 to 2, and we compare our aNNLO and aN$^3$LO 
QCD results with the NNLO QCD results in \cite{CFM} using no expansion in 
$\alpha_S$ (i.e. using the analog of Eq. (\ref{AFBnoexp}) without electroweak 
corrections). We also compare our aN$^3$LO QCD results with data from CDF 
\cite{CDF} and D0 \cite{D0}.  

\begin{figure}
\begin{center}
\includegraphics[width=10cm]{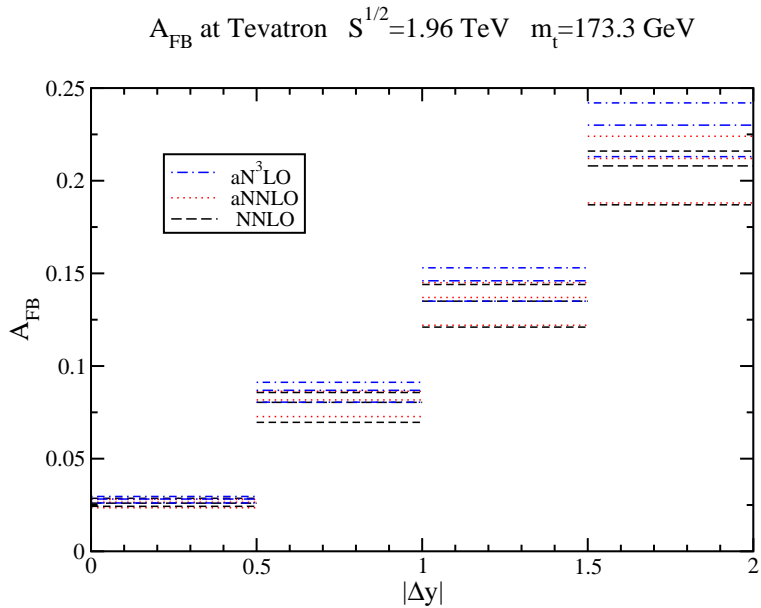}
\caption{The top-quark forward-backward asymmetry at NNLO, aNNLO, and aN$^3$LO 
QCD (without EW corrections) as a function of $|\Delta y|$ in $t{\bar t}$ 
production at the Tevatron with $\sqrt{S}=1.96$ TeV. 
The central lines at each order are with 
$\mu=m_t$, and the other lines display the upper and lower values from scale 
variation over $m_t/2 \le \mu \le 2m_t$.}
\label{AFBplot}
\end{center}
\end{figure}

In Fig. \ref{AFBplot} we show our aNNLO and aN$^3$LO QCD results in the various
$|\Delta y|$ bins and compare them with the NNLO results from \cite{CFM}. 
It is clear that the aNNLO results are very close to the 
exact NNLO results in both central value and scale uncertainty in all bins.
This observation is of course fully consistent with the results for the total 
asymmetry. We also note that the 
aN$^3$LO results enhance the overall values and decrease the theoretical 
uncertainty in all $|\Delta y|$ bins.

\begin{figure}
\begin{center}
\includegraphics[width=10cm]{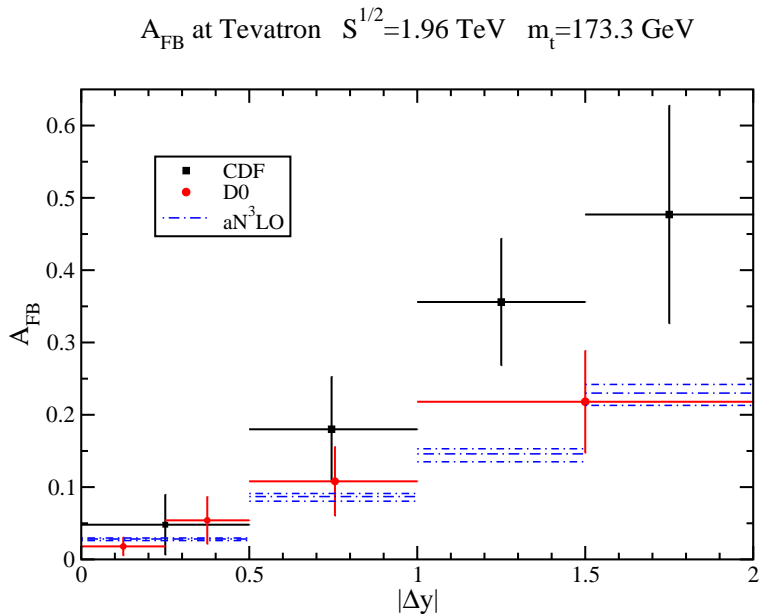}
\caption{The top-quark forward-backward asymmetry at aN$^3$LO QCD 
(without EW corrections)
as a function of $|\Delta y|$ and comparison with CDF \cite{CDF} 
and D0 \cite{D0} data in $t{\bar t}$ production 
at the Tevatron with $\sqrt{S}=1.96$ TeV. 
The central aN$^3$LO lines are with $\mu=m_t$, and the other lines 
display the upper and lower values from scale variation over 
$m_t/2 \le \mu \le 2m_t$.}
\label{AFBexpplot}
\end{center}
\end{figure}

In Fig. \ref{AFBexpplot} we show the comparison of our aN$^3$LO QCD results 
with CDF \cite{CDF} and D0 \cite{D0} data in various $|\Delta y|$ bins.  
The aN$^3$LO results describe the data better than at NNLO. 
Our results are fully consistent with D0 data in all bins; they are 
also in agreement with the CDF data in the first two bins and are within 
two standard deviations or better with CDF data in the highest two bins.

\mysection{Conclusions}

The aN$^3$LO top-quark forward-backward asymmetry at the Tevatron has been 
calculated. It has been shown that the asymmetry is dominated by soft-gluon 
corrections. In fact, at both NLO and NNLO the soft-gluon contribution closely
approximates the complete QCD result. 
Results for $A_{FB}$ have been presented in both the laboratory frame and 
the $t{\bar t}$ rest frame, with and without including electroweak corrections, 
and using two different methods, expansion or no expansion in the strong 
coupling.  
The aN$^3$LO soft-gluon corrections from NNLL resummation substantially enhance 
previous NNLO results 
and they reduce the theoretical uncertainty. The best estimate for the 
asymmetry at aN$^3$LO in the $t{\bar t}$ rest frame is ($10.0 \pm 0.6$)\% 
which is in excellent agreement with experimental results using the full 
Run II data at the Tevatron \cite{CDF,D0}.  
This result for $A_{FB}$ complements the recent aN$^3$LO calculations 
\cite{NKaNNNLO,NKpty} for top-antitop pair production which represent 
the current state-of-the-art in
higher-order calculations for this process.
 
\mysection*{Acknowledgements}
This material is based upon work supported by the National Science Foundation 
under Grant No. PHY 1212472.

\end{document}